\documentclass[pre,twocolumn,superscriptaddress]{revtex4-2}

\usepackage{amsmath,amsthm}    
\usepackage[dvipdfmx]{graphicx}   
\usepackage{verbatim}   
\usepackage{color}      
\usepackage{subfigure}  
\usepackage{hyperref}   
\usepackage[psamsfonts]{amssymb}
\usepackage{amsfonts}
\usepackage{unit}
\usepackage{cases}
\usepackage{here}
\usepackage{bbm}

\newcommand{\figref}[1]{Fig.~\ref{#1}}

\newcommand{\tabref}[1]{Tab.~\ref{#1}}
\newcommand{\secref}[1]{Sec.~\ref{#1}}

\renewcommand{\eqref}[1]{Eq.~(\ref{#1})}
\newcommand{\eeqn}{\end{eqnarray}}

\begin{document}

\title{Characterizing scale dependence of effective diffusion driven by fluid flows}
\author{Yohei Kono}
\affiliation{Department of Electrical Engineering, Kyoto University, Kyoto, Japan}
\author{Yoshihiko Susuki}
\affiliation{Department of Electrical and Information Systems, Osaka Prefecture University, Sakai, Japan}
\author{Takashi Hikihara}
\affiliation{Department of Electrical Engineering, Kyoto University, Kyoto, Japan}


\begin{abstract}
We study the scale dependence of effective diffusion of fluid tracers, specifically, its dependence on the P\'{e}clet number, a dimensionless parameter of the ratio between advection and molecular diffusion.
Here, we address the case that 
length and time scales on which the effective diffusion can be described are not separated from those of advection and molecular diffusion.
For this, we propose a new method for characterizing the effective diffusivity without relying on the scale separation. 
For a given spatial domain inside which the effective diffusion can emerge,
a time constant related to the diffusion is identified by considering the spatio-temporal evolution of a test advection-diffusion equation,
where its initial condition is set at a pulse function. 
Then, the value of effective diffusivity is identified by minimizing the $L_\infty$ distance
between solutions of the above test equation and the diffusion one with mean drift.
With this method, for time-independent gyre and time-periodic shear flows,
we numerically show the scale dependence of the effective diffusivity
and its discrepancy from the classical limits that were derived on the assumption of the scale separation.
The kinematic origins of the discrepancy are revealed as the development of the molecular diffusion across flow cells of the gyre
and as the suppression of the drift motion due to a temporal oscillation in the shear.
\end{abstract}

\keywords{Advection-diffusion, effective diffusion, P\'{e}clet number, scaling law, transition}

\maketitle

\section{Introduction}
\label{sec:intro}

Effective diffusion is a phenomenological concept for describing mixing and dispersion of fluid tracers
(e.g. temperature and chemicals) driven by fluid flows
\cite{csanady2012turbulent,pope_turbulence,homogenization_porous,knauss2016introduction}.
In this paper, we study the scale dependence of the effective diffusion with a new formulation and numerical simulations of rudimentary flow models.

We briefly introduce the concept of effective diffusion as follows.
Let $\mathbb{X}\subseteq\mathbb{R}^n$ ($n=2,3$) be configuration space,
$\vct{x}\in\mathbb{X}$ be location, and $t\geq 0$ be time.
Generally, the concentration profile $\theta(\vct{x},t)$ of fluid tracers
at $\vct{x}$ and $t$ is governed by the following advection-diffusion equation:
\begin{align}    
 \partial_t \theta(\vct{x},t) + \vct{u}(\vct{x},t)\cdot\vct{\nabla}\theta(\vct{x},t)
 = D \Delta \theta(\vct{x},t),
 \label{eq:adv_def_eq}
\end{align}
where $\vct{u}(\vct{x},t)$ represents a pre-defined velocity field on $\mathbb{R}^n$ and satisfies the incompressibility condition. 
The constant $D$ represents the molecular diffusivity of a medium,
$\partial_t$ the differential operator in time, 
$\vct{\nabla}$ and $\Delta$ the vector differential and Laplace operators on $\mathbb{R}^n$.
Following Refs.\,\cite{multiscale_method,bensoussan2011asymptotic},
if the mean-squared displacement of fluid parcels asymptotically increases with $t^2$,
then their macroscopic dispersion can be described by the simple diffusion equation
\begin{align}             
 \partial_t \bar{\theta}(\vct{x},t) = D\sub{eff} \Delta \bar{\theta}(\vct{x},t),
 \label{eq:Deff_eq}
\end{align}
where we call the new concentration profile (function) $\bar{\theta}(\vct{x},t)$
the \emph{up-scaled} field, and $D\sub{eff}$ is known as the \emph{effective diffusivity}.
The meaning of \emph{up-scaling} in this paper is to determine
a pair of finite-volume, connected domain $\Omega \subset \mathbb{X}$
and time-interval $\mathcal{I}:= [0, \tau],~\tau>0$ 
for which the macroscopic dispersion of \eqref{eq:adv_def_eq} is dominant in some sense.

The concept of effective diffusion plays an important role in understanding
wide ranges of physical and engineered systems: see, 
e.g., Refs.\,\cite{soward1987fast,gerz1998transport,mu2008determination}.
Especially, it is of technological importance in analysis and design of thermal dynamics in office buildings.
These dynamics appear on a wide range of scales in both space and time.
They are closely related to the existence of human occupants in a room, 
which work as mobile heat sources that generate buoyancy and as obstacle objects to the air flow.
Consequently, the air flow and the heat flow induced by it develop on the scales of seconds to hours \cite{Kono_JBPS2018}. 
This motivates the use of effective diffusion for modeling the heat phenomenon as shown in Ref.\,\cite{Kono_JBPS2018} by the authors. 
Another motivation is from the modeling of heat transfer inside a building atrium \cite{Kono_SICE2017_diff_english,kono2020modeling}.

Effective diffusion has been characterized in e.g. Ref.\,\cite{Deff_steady} by the so-called P\'{e}clet number 
$Pe := UL/D$: 
the ratio between advection and molecular diffusion,
where $U$ and $L$ are the characteristic velocity and length of the field $\vct{u}(\vct{x},t)$.
The dependence of effective diffusivity on $Pe$, known as a {\em scaling law},
has been studied by several groups of researchers, as reviewed in the following three paragraphs. 

For the steady, two-dimensional periodic gyre flow, 
the effective diffusivity $D\sub{eff}$ exhibits different types of the $Pe$ dependence.  
The $Pe$ dependence is theoretically established in cases of sufficiently small and large $Pe$
\cite{Deff_Horsthmke_first,moffatt1983transport,Deff_steady,isichenko1992percolation,Deff_Marco,Deff_periodic}.
This is exactly expressed as $D\sub{eff} \propto D Pe^2$ for a sufficiently small $Pe$
\cite{Deff_Horsthmke_first,moffatt1983transport,Deff_steady}.
The dependence also behaves like $D\sub{eff} \propto D\sqrt{Pe}$ for a sufficiently large $Pe$
\cite{isichenko1992percolation,Deff_Marco,Deff_periodic},
where the $\sqrt{Pe}$ dependence is explained by the slow diffusive motion of fluid parcels from one periodic flow cell to another.
Moreover, the dependence in the large $Pe$ can be enhanced from $\mathcal{O}(\sqrt{Pe})$ onto $\mathcal{O}(Pe^2)$ 
by adding a mean drift or a time-periodic perturbation
to the steady gyre flow \cite{Deff_Marco,majda1993effect,Deff_periodic,Mezic_SIAM1996},
which is called the \emph{maximal diffusivity}.
In addition to the above limiting cases,
the $Pe$ dependence has been studied numerically for a finite magnitude of $Pe$.
The authors of Ref.\,\cite{Deff_steady} numerically computed the effective diffusivity $D\sub{eff}$ without mean drift
and showed that its $Pe$ dependence is expressed as an arc-like curve,
well fit with both the $\mathcal{O}(Pe^2)$ and $\mathcal{O}(\sqrt{Pe})$ scaling laws
for sufficiently small and large $Pe$, respectively.
The authors of Refs.\,\cite{majda1993effect,majda1999simplified} numerically showed that
by adding a mean drift to the gyre,
the $Pe$ dependence can exhibit an exponent lower than $\mathcal{O}(\sqrt{Pe})$ at a finite $Pe$, 
referred to as the {\em crossover effect}.
Although these numerical studies contain significant progress,
their kinematic origins are not necessarily clarified.

For time-invariant and time-periodic shear flows, 
the effective diffusivity $D\sub{eff}$ has been calculated as closed-form functions of $Pe$ without limitation of the magnitude of $Pe$ 
\cite{zel1982exact,frankel1989foundations,isichenko1992percolation}.
Its kinematic origin is stated in \cite{majda1999simplified} as the combination of the molecular diffusion between streamlines
and of the drift motion along the shear. 
The $D\sub{eff}$ can take $\mathcal{O}(Pe^2)$ only for the time-invariant shear \cite{Mezic_SIAM1996}.
Also, by analytical calculation of the time-dependent, finite-scale dispersion induced by the shear flow,
the authors of Ref.~\cite{lin2010models} 
proposed different formulas of mixing efficacy of fluid parcels
so as to explain distinct $Pe$ scalings of the effective diffusivity.

Most of the above theoretical and numerical studies are conducted with the so-called homogenization limit \cite{multiscale_method,bensoussan2011asymptotic},
where the original length-scale (or time-scale) of mixing and dispersion characterized by $U$, $L$, and $D$ is
clearly separated from the length $L_\Omega$ of the domain $\Omega$
(or the time constant $\tau$ of the interval $\cal I$). 
The assumed scale separation is crucial to deriving the above explicit formulae of the $Pe$ dependence of effective diffusion.
The author of Ref.\,\cite{fannjiang2002time} has found finite time-scales on which
a limit theorem of the homogenization holds for sufficiently large $Pe$.
Also, recent studies \cite{owhadi2007metric,owhadi2008homogenization,berlyand2010flux,babuska2011optimal,constantin2008diffusion}
have elaborated on analytical and numerical frameworks for computing the effective diffusivity where no scale separation is assumed.
However, these studies are conducted without reference to the kinematic origin of the $Pe$ dependence. 

To the best of our survey,
the kinematic study on the $Pe$ dependence of effective diffusion relies on the scale separation,
that is, it has not been explored
how the effective diffusion is affected by wide ranges of the magnitudes of $L_\Omega$ and $\tau$.
As mentioned above,
the authors of Ref.\,\cite{majda1999simplified} explained the effective diffusion for the shear flows
by assuming that $\tau$ is so large that a fluid parcel can transit from one streamline of the shear to another by molecular diffusion. 
Also, in Ref.~\cite{lin2010models}, the mixing efficacy was defined
by integrating a fundamental solution of \eqref{eq:adv_def_eq} over the infinite time interval.
Thus, the shear-induced transport for a finite $\tau$ has not been considered in the context of effective diffusion.
It should be emphasized that
the need of investigating the effective diffusion for such a finite $\tau$ is pointed out in simulation studies on oceanography and atmospheric science
\cite{shuckburgh2009robustness,shuckburgh2009understanding,d2009local}
and motivated by the engineering of thermal dynamics in office buildings as mentioned above.

The purpose of this paper is to characterize the scale dependence of effective diffusivity $D\sub{eff}$
over the range of scale on which the previous theories and methods for illustrating the effective diffusion were constructed.
More specifically, we numerically investigate
how the $Pe$ dependence of $D\sub{eff}$ is affected by the parameters $L_\Omega$ and $\tau$ with finite magnitudes.
To this end, as the first part,
we determine the values of $L_\Omega$ and $\tau$
for which the macroscopic dispersion of \eqref{eq:adv_def_eq} is {\em dominant} in some sense.
In this paper, we will show that the macroscopic dispersion is 
dominant for a given $L_\Omega$ 
if $\tau$ takes the same order as the time when it takes for a fluid parcel to travel over $\Omega$.
Then, as the second part, we focus on the two rudimentary models---time-independent gyre and time-periodic shear flows---and show computational results on their effective diffusivity $D\sub{eff}$ so that its $Pe$ dependence is numerically described. 
The computation scheme is formulated and conducted in an engineering framework.

The contributions of this paper are twofold.
First, we develop a new method for identifying the effective diffusion 
without relying on the scale separation.
For given $\Omega$ with $L_\Omega$, using techniques from control and optimization,
we propose to identify the time-scale $\tau$ and then the effective diffusivity $D\sub{eff}$.
Here, $\tau$ is identified by considering the spatio-temporal evolution of a test advection-diffusion equation,
where its initial field is set at a pulse function, whose definition is presented in \secref{sec:adv_diff_effective_diff}.
This is analogous to the identification of dynamic responses of
linear time-invariant systems using an impulsive input \cite{ljung1999system}.
The identification of $\tau$ is novel especially for a finite magnitude of $Pe$,
where advection and molecular diffusion are comparable so that
the identification based on vanishing molecular diffusion \cite{Deff_periodic,fannjiang2002time}
is not available.
Note that the authors of Ref.~\cite{artale1997dispersion} introduced the shortest time reaching a certain predetermined radius,
which is the boundary of $\Omega$ in this paper,
so as to determine finite-size Lyapunov exponent and dispersion.
The connection between our identification and the above determination is shown in \secref{sec:adv_diff_effective_diff}.
Using the time-scale $\tau$,
$D\sub{eff}$ is identified by minimizing the $L_\infty$ distance
between solutions of the above test equation and the diffusion
one (\ref{eq:Deff_eq}) with mean drift. 
Second, for the two rudimentary models, 
we reveal the kinematic origins of the $Pe$ dependence of $D\sub{eff}$, 
where no scale separation is assumed. 
For the gyre flow,
we numerically show that the $Pe$ dependence can change from $\mathcal{O}(\sqrt{Pe})$ at a finite $Pe$.
Based on a finite magnitude of $\tau$,
we newly estimate the length of finite-time dispersion of fluid parcels due to the effective diffusion,
by which the change of the $Pe$ dependence is explained
as the development of molecular diffusion between flow cells of the gyre.  
For the shear flow, we numerically show that a finite magnitude of $\tau$ causes the deviation of $D\sub{eff}$ from the closed-form function of $Pe$.
The deviation is explained as the degree of insufficiency of molecular diffusion between streamlines.

The rest of this paper is organized as follows.
\secref{sec:adv_diff_effective_diff} is devoted to developing the method for characterizing the effective diffusion
where no scale separation is assumed.
\secref{sec:gyre_results} shows computational results of the effective diffusion for the gyre flow.
By sweeping the molecular diffusivity $D$, 
we will validate our method with the $\sqrt{Pe}$ scaling
and show its breakdown when no scale separation holds.
In \secref{sec:shear_results}, we study the $Pe$ dependence of the effective diffusion for the shear flow,
where the time-constant of molecular diffusion between streamlines
is on the same order as drift motion by the shear.
\secref{sec:conclusion} is the conclusion with brief summary and discussion on generality of the proposed method and the physical findings.

\section{Pulse-based method for characterizing effective diffusion}
\label{sec:adv_diff_effective_diff}

This section is devoted to developing a method for characterizing the effective diffusion without assumption of scale separation.
Here, for a given $\Omega \in \mathbb{X}$,
the time-scale $\tau$, effective diffusivity $D\sub{eff}$, and associated error term
as the result of the approximation of effective diffusion are determined
via the spatio-temporal evolution of a pulse function,
which is defined in the next paragraph.
Below, we assume that $\vct{u}(\vct{x},t)$ is periodic in both space and time. 
Precisely, it is assumed that the signal $\vct{u}(\vct{x},t)$ for any fixed $\vct{x}\in\Omega$ has a finite number of peaks in Fourier spectrum, for which we use $\tau_0$ to represent the fundamental period.
It is also assumed that $\tau_0$ and the fundamental period $L$ of $\vct{u}(\vct{x},t)$ in space are known a priori,
and that $L$ is smaller than $L_\Omega$.
These assumptions hold for our rudimentary models in this paper.

Let us introduce a test equation used for this method. 
Recalling that effective diffusivity is not sensitive to an initial field $\theta_0$
(see, e.g., Ref.\,\cite{shuckburgh2009robustness})
and, in certain cases, can be a functional of $\vct{u}(\vct{x},t)$ \cite{multiscale_method},
we evaluate it via the following test partial differential equation (PDE):
\begin{align}                                                                                 
 \partial_t \rho(\vct{x},t) + \vct{u}(\vct{x},t)\cdot\vct{\nabla}\rho(\vct{x},t)
 &= D \Delta \rho(\vct{x},t).
 \label{eq:test_adv_diff}
\end{align}
Importantly, the initial field $\rho(\vct{x},0)$ is fixed at a certain class of functions that we call the ``pulse'' function $\rho_0(\vct{x})$. 
For our method, 
$\rho_0(\vct{x})$ should be taken as a function such that it is supported in the interior of $\Omega$ and localized (in $\vct{x}$) in terms of $L_\Omega$ so that the diffusion phenomenon clearly develops in space. 
Regarding this, there are multiple choices of the pulse function;
e.g., the Dirac's delta function can be used in context of mathematical analysis.
This choice is used in terms of the Lagrangian approach \cite{mezic1994dynamical,Mezic_SIAM1996,lin2010models,majda1999simplified},
where effective diffusion is described via the long-term evolution of the density of fluid parcels
that start from the support of the delta function.
In this paper, in order to gain better regularity of the problem for numerics,
we use the Gauss function as $\rho_0(\vct{x})$ and control its length scale by the variance parameter $\sigma$: 
\begin{align}
 \rho_0(\vct{x})=\exp{\left(\frac{\|\vct{x}-\vct{c}_\Omega\|^2}{\sigma^2}\right)}
 \label{eq:gaussian_pulse},
\end{align}
where $\vct{c}_\Omega$ stands for a geometric center (centroid) 
of $\Omega$, and $\|\cdot\|$ for the vector norm.
To clearly investigate the effect of advection with its spatial period $L$, 
we fix $\sigma$ such that its order of magnitude is equal to and smaller than that of $L$.

As the first step of the method, for given $\Omega$,
we determine the time-scale $\tau$ relevant to the dispersion of fluid parcels in $\Omega$.
The $\tau$ is identified via the parameter $\tau_{\Omega,\alpha}$ that is a function of
$\Omega$ and a small parameter $\alpha$ for judging if a fluid parcel reaches a given position or not. 
Let $\partial\Omega$ be the boundary curve or surface of $\Omega$.
For given $\vct{x}\in\Omega$,
if a fluid parcel reaches $\vct{x}$ from an initial position close to $\vct{c}_\Omega$,
then there exists an onset time $t$, denoted by $\tilde{\tau}_{\Omega,\alpha}(\vct{x})$, such that 
$\rho(\vct{x},t)=\alpha \int_{\Omega}\rho_0(\vct{y})\mu(\dd\vct{y})/|\Omega|$
holds, where $\mu({}_\sqcup)$ is a standard measure on $\Omega$,
and $|\Omega|=\int_\Omega\mu(\dd\vct{y})$ coincides with the volume or area of $\Omega$.
Here, by supposing that
a fluid parcel can reach the boundary $\partial\Omega$
by the advection and diffusion,
it is possible to estimate the time $\tau_{\Omega,\alpha}$ given by
\begin{align}
 \tau_{\Omega,\alpha}=\inf_{\vct{x}\in\partial\Omega}\tilde{\tau}_{\Omega,\alpha}(\vct{x}).
 \label{eq:interval_def}
\end{align}
This is an approximation of the first time when the pulse of \eqref{eq:gaussian_pulse} hits $\partial\Omega$.

Here, we comment on how to determine the parameter $\alpha$.
As $\alpha$ becomes large, it is possible to clearly detect the hitting of the pulse;
however, it requires long time for the computation of \eqref{eq:interval_def}.
Also, for avoiding the trivial case $\tau_{\Omega,\alpha}=0$, 
the initial $\rho_0(\vct{x})$ should be smaller than the threshold
$\alpha\int_{\Omega}\rho_0(\vct{y})\mu(\dd\vct{y})/|\Omega|$
at every $\vct{x}\in\partial\Omega$.
Thus, $\alpha$ needs to satisfy the following inequality: 
\begin{align*}
\alpha > \frac{\displaystyle |\Omega|
\sup_{\vct{x}\in\partial\Omega}\rho_0(\vct{x})}
{\int_{\Omega}\rho_0(\vct{y})\mu(\dd\vct{y})} 
=: \Lambda_{\Omega,\sigma}.
\end{align*}
Below, we will fix $\alpha$ at a small value satisfying the above inequality,
and hence $\tau_{\Omega,\alpha}$ can be computed in practical time. 
Regarding this, we will also show the $L_\Omega$ dependence of $\Lambda_{\Omega,\sigma}$
(see \figref{fig:detection_threshold}).

Next, we identify the drift-oriented transport of $\rho$ during the interval $[0,\tau]$
in order to eliminate it for estimating the effective diffusivity. 
Inspired by the averaging method \cite{multiscale_method},
we quantify the so-called bulk movement of fluid parcels by 
\begin{align}
 \vct{c}(t) := \frac{1}{C_0}
 \int_{\Omega} \vct{x}\rho(\vct{x},t) \mu(\dd\vct{x}),\,\,\,t\in[0,\tau]
 \label{eq:bulk_movement_def},
\end{align}
where $C_0 := \int_{\Omega} \rho_0(\vct{y})\mu(\dd\vct{y})$. 
The meaning of $\vct{c}(t)$ is described below in terms of the averaging method. 
As in Ref.\,\cite{multiscale_method} we suppose that $\Omega$ is point-symmetric with respect to its center $\vct{c}_\Omega$.
Then, by combining this with the periodicity of $\vct{u}$, the equality
$\int_{\partial\Omega}\vct{x}\{(\rho\vct{u})\cdot\vct{n}\}\mu(\dd\vct{x})= \vct{0}$ holds,   
where $\vct{n}(\vct{x})$ is the normal vector at point $\vct{x}$ on $\partial\Omega$.
Also, when the pulse of \eqref{eq:gaussian_pulse} does not hit $\partial\Omega$ at $t < \tau$,
the gradient $\vct{\nabla}\rho(\vct{x},t)$ is negligible at any $\vct{x}\in\partial\Omega$. 
Then, the time derivative of $\vct{c}(t)$ is derived as
\begin{align}
 \frac{\dd \vct{c}}{\dd t}
 =&\frac{1}{C_0} \int_{\Omega} \vct{x}
 \left\{
 -\vct{\nabla}\cdot
 {(\rho\vct{u})}
 + D\Delta\rho
 \right\}
 \mu(\dd\vct{x}),
 \nonumber\\
 =& -\frac{1}{C_0}\int_{\partial\Omega}
 \vct{x}\{(\rho\vct{u})\cdot\vct{n}\}\mu(\dd\vct{x})\nonumber\\
 &+\frac{1}{C_0}\int_\Omega(\rho\vct{u})\mu(\dd\vct{x})
 +\frac{D}{C_0}\int_{\partial\Omega}(\vct{\nabla}\rho)\cdot\vct{n}\mu(\dd\vct{x}),
 \nonumber\\
 \sim
 &\frac{1}{C_0}\int_{\Omega} \vct{u}(\vct{x},t)\rho(\vct{x},t)\mu(\dd\vct{x}),
 \label{eq:effecti_velocity_extension}
\end{align}
where we use the integration by parts for each element
to move from the first line to the second.
\eqref{eq:effecti_velocity_extension} corresponds to the classical effective velocity \cite{multiscale_method} 
if $\rho(\vct{x},t)/C_0$ is regarded as a probability density function on $\Omega$. 
This clearly shows that $\vct{c}(t)$ in \eqref{eq:bulk_movement_def} represents the averaged (in space) movement of fluid parcels in $\Omega$.
Since the effective velocity in \cite{multiscale_method}
represents the spatio-temporal mean of $\vct{u}(\vct{x},t)$, we define 
\begin{align}
 \bar{\vct{U}}_{\Omega,\alpha} := \frac{\vct{c}(\tau_{\Omega,\alpha})-\vct{c}_\Omega}
 {\tau_{\Omega,\alpha}},
 \label{eq:effective_velocity}
\end{align}
as the {\em effective velocity} to describe the drift transport over $[0,\tau]$.

Finally, we develop the concrete step of identifying the effective diffusivity.
The key idea of this is to approximate the local dispersion of fluid parcels from the center $\vct{c}_\Omega$ as the effective diffusion. 
This approximation is conducted via the following diffusion equation that allows the mean flow $\bar{\vct{U}}_{\Omega,\alpha}$:
\begin{align}
 \left(
 \partial_t + \bar{\vct{U}}_{\Omega,\alpha}\cdot\vct{\nabla}
 \right)
 \hat{\rho}_{\bar{D},\Omega,\alpha}(\vct{x},t)
 &= \bar{D} \Delta \hat{\rho}_{\bar{D},\Omega,\alpha}(\vct{x},t),
 \label{eq:test_meanflow_diff}
\end{align}
where $\bar{D}$ is a candidate of the effective diffusivity,
and $\hat{\rho}_{\bar{D},\Omega,\alpha}(\vct{x},t)$ represents the concentration profile of fluid parcels driven by the mean flow and the diffusion with its coefficient $\bar{D}$.
Here, since the length-scale $L_\Omega$ of the target domain is larger than that of $\rho_0(\vct{x})$,
it is natural to understand the dispersion from $\vct{c}_\Omega$ via a solution of an isotropic homogeneous diffusion equation on $\mathbb{R}^n$.
Its fundamental solution (i.e. heat kernel) is given by a time-varying Gaussian function \cite{Evans2010partial}.
In this, the dispersion is represented as the temporal change of the ``height'' (maximum value) of the kernel.
The height of the kernel is well captured by its $L_\infty$ norm rather than $L_2$ norm.
Thus, based on the $L_\infty$ distance between solutions of advection-diffusion and effective diffusion equations in Ref.\,\cite{multiscale_method},
we introduce the $L_\infty$ distance to identify the effective diffusivity as follows: 
\begin{align}
 \hat{d}_{\bar{D},\Omega,\alpha}(t):=
 \sup_{\vct{x}\in\Omega}
 |\rho(\vct{x},t) - \hat{\rho}_{\bar{D},\Omega,\alpha}(\vct{x},t)|.
 \label{eq:dist_series}
\end{align}
This $\hat{d}_{\bar{D},\Omega,\alpha}$ can change in not only $\bar{D},\Omega$ but also $t$.
In particular, it can have a wide range of time-frequency spectrum
beyond that of $\hat{\rho}_{\bar{D},\Omega,\alpha}(\vct{x},t)$
since $\rho(\vct{x},t)$ is affected by the time-dependent $\vct{u}(\vct{x},t)$
but $\hat{\rho}_{\bar{D},\Omega,\alpha}(\vct{x},t)$ by the constant $\bar{\vct{U}}_{\Omega,\alpha}$. 
However, by the original notion of effective diffusion,
such high-frequency components should be filtered out
for the modeling of macroscopic mixing and dispersion.
Thus, for estimating better $\bar{D}$ in the $L_\infty$ sense,
it is necessary to filter out high-frequency components
whose time-scale is smaller than a pre-defined constant denoted as $\tau_0$. 
There exist many methods for this low-pass filtering in signal-processing textbooks:
see e.g., Ref.\,\cite{smith1997scientist}.  
In this paper, for simplicity of implementation,
we use the first-order filter to derive
a \emph{smoothed} error $d_{\bar{D},\Omega,\alpha}(t)$ from $\hat{d}_{\bar{D},\Omega,\alpha}(t)$ as follows:
\begin{equation}
 \left(\tau_0 \frac{\dd}{\dd t} + 1 \right) 
 d_{\bar{D},\Omega,\alpha}(t) = \hat{d}_{\bar{D},\Omega,\alpha}(t),
 \label{eq:dist_LPF}
\end{equation}
where the smoothed error $d_{\bar{D},\Omega,\alpha}(t)$ is initialized as
$d_{\bar{D},\Omega,\alpha}(0) = \hat{d}_{\bar{D},\Omega,\alpha}(0)$.
With this, we search a value of $\bar{D}$
that minimizes the cost function defined by
\begin{align}
 \sup_{t\in\mathcal{I}} d_{\bar{D},\Omega,\alpha}(t),
 \label{eq:Deff_def}
\end{align}
where its minimizer is referred to as $\bar{D}_{\Omega,\alpha}$,
corresponding to an estimated value of the effective diffusivity $D\sub{eff}$ for given $\Omega$ and $\alpha$.

\begin{figure*}[!tb]
\begin{center}
  \includegraphics[width=0.8\hsize]{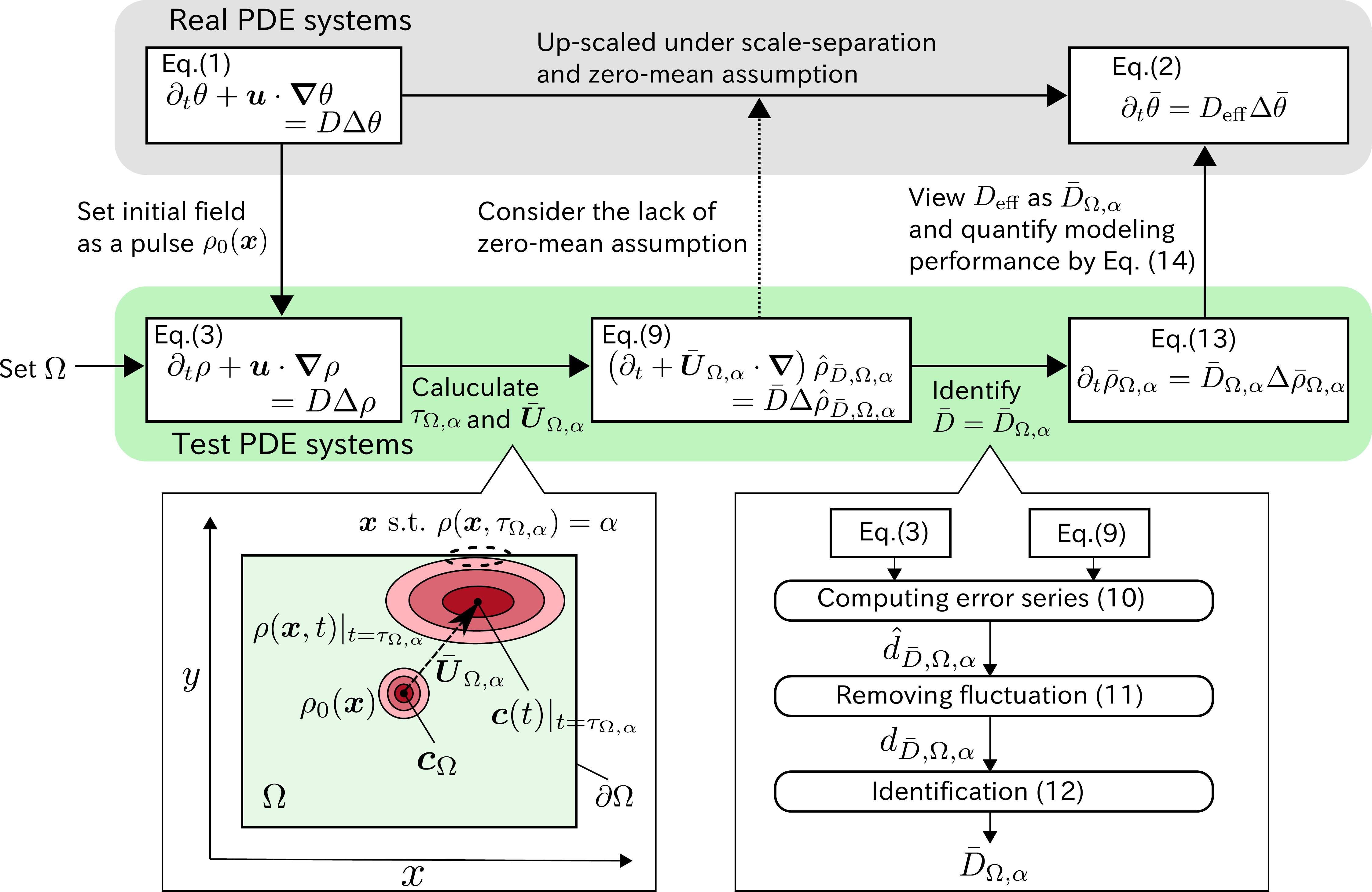}
  \caption{Schematic diagram of the proposed characterization of effective diffusion.}
  \label{fig:PDE_diagram}
\end{center}
\end{figure*}

Moreover, we introduce a metric for the effective diffusion to investigate its performance and application limit,
which we will compare with an error metric given by the homogenization in order to validate the proposed method.
Recalling that the goal of the homogenization in Ref.~\cite{multiscale_method,bensoussan2011asymptotic} is to describe macroscopic dispersion of fluid parcels as the pure diffusion,
we compare \eqref{eq:test_adv_diff} with the following diffusion equation:
\begin{align}
 \partial_t \bar{\rho}_{\Omega,\alpha}(\vct{x},t) = \bar{D}_{\Omega,\alpha} \Delta \bar{\rho}_{\Omega,\alpha}(\vct{x},t),
 \label{eq:test_diff}
\end{align}
where $\bar{\rho}_{\Omega,\alpha}(\vct{x},t)$ represents the concentration profile of fluid tracers driven by the pure diffusion with its coefficient $\bar{D}_{\Omega,\alpha}$.
For this, we consider the difference between the two initial fields:
$\theta_0(\vct{x})$ assumed by the classical homogenization and $\rho_0(\vct{x})$ by our method.
In \cite{multiscale_method}, $\theta_0(\vct{x})$ is assumed to be a periodic function with its period sufficiently smaller than $L_\Omega$, implying that fluid parcels are homogeneously located on $\Omega$.
However, in this paper, the initial field is set at the pulse function $\rho_0(\vct{x})$
so that fluid parcels stay near its center $\vct{c}_\Omega$ for small $t$. 
The local dispersion does not appear in the homogenization approach and therefore should be excluded
for investigating the performance of the effective diffusion.
By taking this into account, it is desirable to introduce a metric
based on the concentration profile after fluid parcels are sufficiently dispersed over $\Omega$.
In this paper, we use the following metric $E_{\Omega,\alpha}$ at the onset time $t = \tau_{\Omega,\alpha}$: 
\begin{align}
 E_{\Omega,\alpha}:=
 \sup_{\vct{x}\in\Omega}
 \left|
 \rho(\vct{x},\tau_{\Omega,\alpha})
 - \bar{\rho}_{\Omega,\alpha}(\vct{x},\tau_{\Omega,\alpha})
 \right|{\color{blue}.}
 \label{eq:distance_def}
\end{align}
The method developed above is summarized as a schematic diagram in \figref{fig:PDE_diagram}.

Here, we discuss the connection between the above identification and the determination of finite-size Lyapunov exponent and dispersion \cite{artale1997dispersion}.
In Ref.~\cite{artale1997dispersion}, the authors introduced the ``doubling time,''
implying the time that it takes for the initial size of a cloud of fluid parcels growing into its double size.
Recalling that \eqref{eq:interval_def} is an approximation of the first time when the pulse of \eqref{eq:gaussian_pulse} hits $\partial \Omega$,
the doubling time can be computed by setting the length scale $\sigma$ of Eq.~(4) at a half of the length $L_\Omega$.
Then, by the dimensional analysis based on the doubling time and the initial size of a cloud,
they determined the finite-size Lyapunov exponent and the finite size diffusion coefficient
so as to characterize dispersion of parcels.
In this, there is no investigation into the possibility of deriving the spatio-temporal evolution $\bar{\theta}(\vct{x},t)$ from \eqref{eq:Deff_eq}. 
This implies the difference from our method,
where $D\sub{eff}$ is directly estimated by minimizing the cost function (\ref{eq:Deff_def})
so that $\bar{\theta}(\vct{x},t)$ can be obtained.

Before the concrete applications, it is valuable to mention the computational aspect of the proposed method.
The optimization poses, unfortunately, a non-convex problem,
and hence its local minimizer does not imply the global one.
In the applications below,
the explicit formulae of rudimentary flow models are available so that
the order of ``true'' effective diffusivity $D\sub{eff}$ can be estimated a priori. 
Restricted to the true order, the cost function (\ref{eq:Deff_def}) likely exhibits the convex property.
Then, we will conduct the grid search algorithm \cite{hsu2003practical,chicco2017ten} 
in order to search the minimizer of \eqref{eq:Deff_def}.
However, when the proposed method is used in the combination with the computational fluid dynamics (CFD) (see, e.g., Ref.\,\cite{pope_turbulence}),
the true order of $D\sub{eff}$ cannot be estimated because of the complex nature of $\vct{u}(\vct{x},t)$. 
In this case, the optimization requires global techniques such as meta-heuristics \cite{talbi2009metaheuristics} for locating a global minimum of \eqref{eq:Deff_def}
while avoiding a local one.

\section{Scale dependence for gyre flow}
\label{sec:gyre_results}

In this section, we apply the proposed method to a simple gyre flow
and characterize its effective diffusion as a function of the molecular diffusivity $D$.
The purposes of this application are two-fold.
The first one is to validate the method by comparison with the traditional theory of $D$-dependence of the effective diffusivity $D\sub{eff}$
(see \eqref{eq:double_gyre_Deff} below),
which is based on the scale separation.
The second one is to show a breakdown of the theory beyond the regime where the scale separation holds,
which we will refer to as the transition of effective diffusion. 

\begin{figure}[!tb]
 \begin{center}
  \includegraphics[width=0.78\hsize]{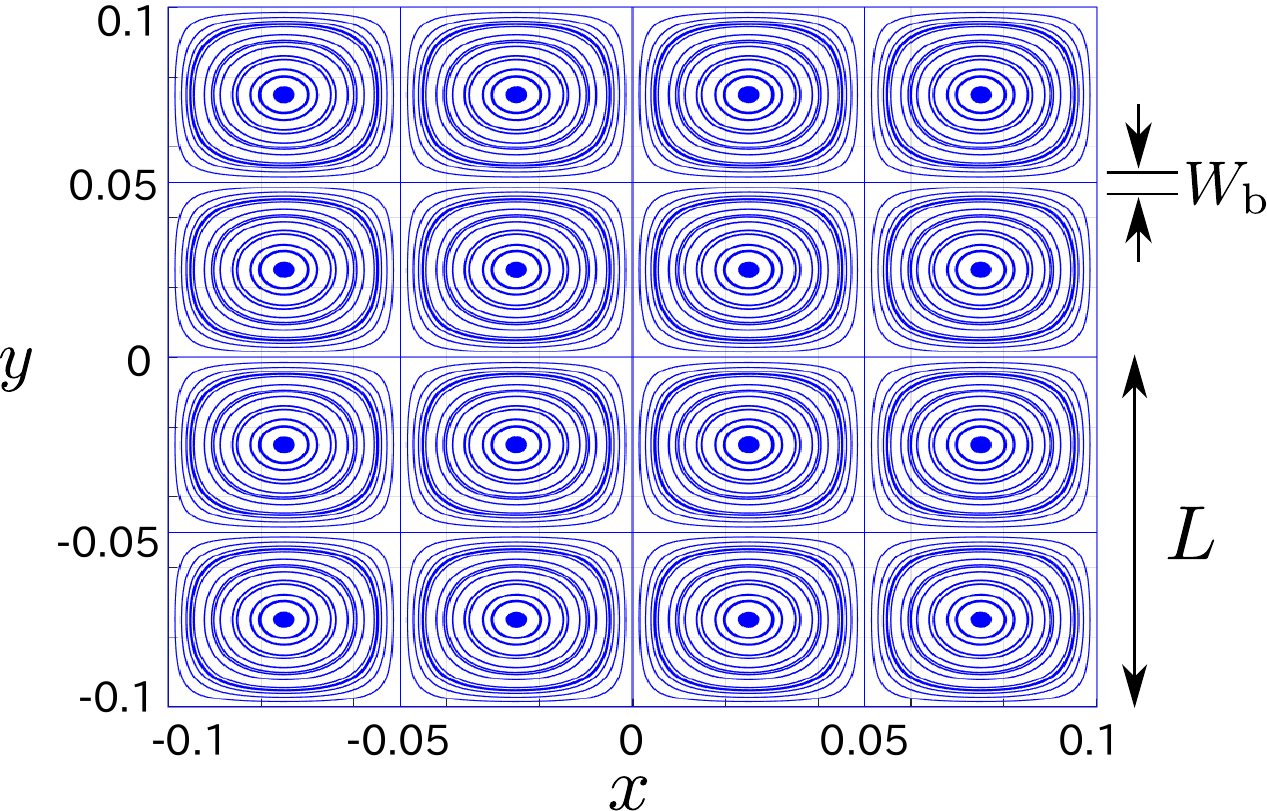}
  \caption{Illustration of streamlines of the vector field (\ref{eq:double_gyre}).}
  \label{fig:gyre_flow}
 \end{center}
\end{figure}
The model flow is time-invariant and represented by the following two-dimensional vector field:
for $\vct{x}=[x\ y]^\top$,
\begin{align}
 \vct{u}(\vct{x}) = \left[
 \begin{array}{c}
  -U\sin(2\pi x/L) \cos(2\pi y/L)\\
  U\cos(2\pi x/L) \sin(2\pi y/L)
 \end{array}
 \right],\label{eq:double_gyre}
\end{align}
where $\mathbb{X} = [-0.5, 0.5]\times[-0.5, 0.5]$,
$U$ and $L$ are the characteristic velocity and the length of the flow field \cite{Note1}.
Figure~\ref{fig:gyre_flow} illustrates streamlines of the vector field (\ref{eq:double_gyre}) with $L=0.1$ 
in the subdomain $[-0.1,\ 0.1]\times[-0.1,\ 0.1]$.
Also, the target domain $\Omega$ is set at a square $[-L_\Omega/2, L_\Omega/2]\times[-L_\Omega/2, L_\Omega/2]$
with the controllable length $L_\Omega$.

Under the scale separation (i.e. $L_\Omega \gg L$),
the scaling law and error convergence of the effective diffusivity for \eqref{eq:double_gyre} are well-known. 
For $Pe := UL/D$, the $\sqrt{D}$ scaling law of $D\sub{eff}$ is given in Refs.~\cite{isichenko1992percolation,Deff_periodic}:
for sufficiently large $Pe$,
\begin{align}
 D\sub{eff} \propto D \sqrt{Pe} = \sqrt{DUL}
 \label{eq:double_gyre_Deff}.
\end{align}
Also, for the scaling ratio $L/L_\Omega$, 
the upper bound of the following error $\tilde{E}_{\Omega,\alpha}$ is given in Ref.\,\cite{multiscale_method} by virtue of the homogenization:
Under the assumption that the initial field $\theta_0(\vct{x})$ of \eqref{eq:adv_def_eq} is periodic,
we have
\begin{align}
 \tilde{E}_{\Omega,\alpha} :=& 
 \sup_{\vct{x}\in\Omega}
 |\theta(\vct{x},\tau_{\Omega,\alpha}) - \bar{\theta}(\vct{x},\tau_{\Omega,\alpha})|\nonumber\\
 \leq& \sup_{\vct{x}\in\Omega,t\in[0,\tau_{\Omega,\alpha}]}
 |\theta(\vct{x},t) - \bar{\theta}(\vct{x},t)|
 < L/L_\Omega
 \label{eq:error_convergence}.
\end{align}
Below, we will validate the proposed method by not only reproducing these results but also exploring their discrepancy from classical limits.

In this paper, it is referred to that the scaling law (\ref{eq:double_gyre_Deff}) becomes irrelevant for explaining the underlying transport phenomenon,
as the transition of effective diffusion.
The occurrence of the transition has been numerically shown in literature, e.g., Refs.\,\cite{majda1993effect,majda1999simplified}.
Here, we point out that it results from the enhancement of the molecular diffusivity $D$. 
The scaling law is originated from the formation of a diffusive boundary layer with width $W\sub{b}$,
implying a small neighborhood of separatrix of the flow \cite{isichenko1992percolation,Deff_periodic};
see the vertical and horizontal lines in \figref{fig:gyre_flow}.
Because advection is dominant in a flow cell,
the two time constants---the time $\tau\sub{a} := L/U$ to go around a cell by advection and the time $\tau\sub{d} := W\sub{b}^2/D$ to traverse diffusively the boundary layer---can be balanced,
i.e., $\tau\sub{a} \simeq \tau\sub{d}$.
Thus, the width $W\sub{b}$ is determined as
\begin{align}
 W\sub{b} = \sqrt{DL/U}. \label{eq:boundary_layer}
\end{align}
Then, $D\sub{eff}$ is estimated by multiplying
a usual random-walk expression $L^2/(L/U)$
by the ratio of particles $W\sub{b}/L$ lying in the boundary layer,
leading to the scaling (\ref{eq:double_gyre_Deff}) in Ref.\,\cite{isichenko1992percolation}.
Here, it naturally follows for large $D$ that
$\tau\sub{a}$ and $\tau\sub{d}$ are not comparable.
Then, the balance between advection and molecular diffusion can break down
and cause the transition of the scaling law.
This will be numerically clarified below.

Let us summarize the current setting of numerical simulations.
We used the parameters $\sigma = 0.04$, $U = 1$, and $L = 0.1$.
\begin{table}[!tb]
\caption{Setting of of numerical simulations for gyre flow}
\label{tab:scale_experiment}
\begin{center}
\begin{tabular}{ccc} \hline\hline
Setting    & $D$ & $L_\Omega$ \\\hline
\#1 & $10^{-4}$ & $\{0.2, 0.3, \ldots, 0.8\}$ \\
\#2 & $\{10^{-6}, 10^{-5.9}, \ldots, 10^{-4}\}$ & 0.7 \\\hline\hline
\end{tabular}
\end{center}
\end{table}
\begin{figure}[!tb]
 \begin{center}
  \includegraphics[width=0.9\hsize]{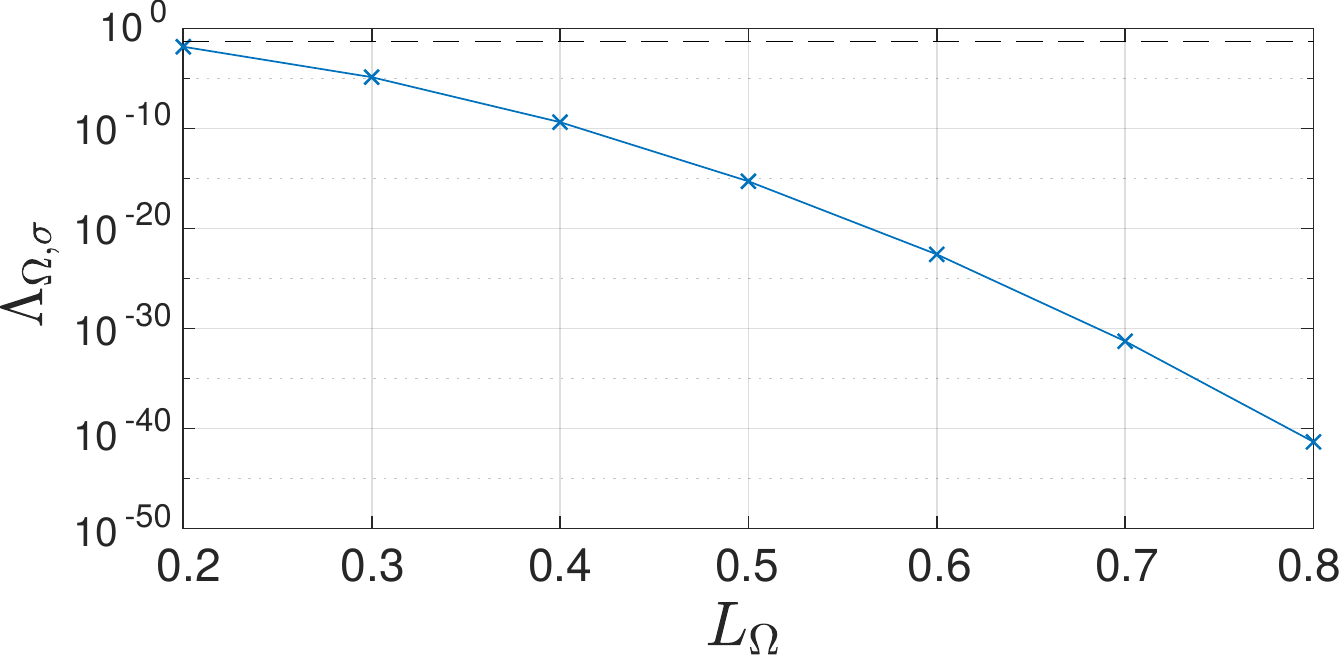}
  \caption{Numerically computed  $\Lambda_{\Omega,\sigma}$ for each $L_\Omega$.
  Here the variance parameter $\sigma$ is fixed at 0.04.
  The blue cross marks denote the calculated values, and the broken line denotes $\Lambda_{\Omega,\sigma} = 0.05$.}
  \label{fig:detection_threshold}
 \end{center}
\end{figure}
Also, as shown in \tabref{tab:scale_experiment},
we varied $L_\Omega$ and $D$ in order to confirm whether or not the scaling law (\ref{eq:double_gyre_Deff})
and the error convergence (\ref{eq:error_convergence}) are reproduced by the proposed method. 
Moreover, to show that the effective diffusion is not sensitive to the selection of initial fields,
we computed the error metric $\tilde{E}_{\Omega,\alpha}$ in \eqref{eq:error_convergence} with a certain initial field $\theta_0$.
Since the effective diffusivity $\bar{D}_{\Omega,\alpha}$ and its time constant $\tau_{\Omega,\alpha}$ are computed by the dispersion from the single center $\vct{c}_\Omega$,
any peak of $\theta_0(\vct{x})$ should not be located at the center $\vct{c}_\Omega$
for distinguishing $E_{\Omega,\alpha}$ from $\tilde{E}_{\Omega,\alpha}$.
Thus, we set $\theta_0(\vct{x})$ at the following mixed-Gaussian distribution:
\begin{align}
 \theta_0(\vct{x}) = \exp\left( \frac{\|\vct{x}-\vct{c}_1\|}{\sigma^2} \right)
 - \exp\left( \frac{\|\vct{x}-\vct{c}_2\|}{\sigma^2} \right),
 \label{eq:initial_field}
\end{align}
where the positions of peaks were denoted by $\vct{c}_1 = [0.4\ 0.4]^\top$ and $\vct{c}_2 = [0.6\ 0.6]^\top$.
With this, we computed $\tilde{E}_{\Omega,\alpha}$ by setting $\tau = \tau_{\Omega,\alpha}$ and $D\sub{eff} = \bar{D}_{\Omega,\alpha}$ in Eqs.~(\ref{eq:adv_def_eq}) and (\ref{eq:Deff_eq}).
Here, by varying $L_\Omega$, we plotted the lower bound of $\alpha$,
namely $\Lambda_{\Omega,\sigma}$ in \figref{fig:detection_threshold}, 
which took its maximum 0.0154 at $L_\Omega = 0.2$.
Thus we fixed $\alpha$ at 0.05, 
which was on the same order as (and larger than) the above maximum,
and by which we could compute $\tau_{\Omega,\alpha}$ in practical time. 
All numerical simulations were conducted by the forward-time centered-space scheme
\cite{press1988numerical}, 
where the discretization steps were set at 0.005 in space and 0.001 in time.
The minimizer of \eqref{eq:Deff_def} was searched by the grid search
with the candidates $\bar{D} \in \{10^{-5},10^{-4.99},\ldots,10^{-2.5}\}$.

\begin{figure}[!tb]
 \begin{center}
  \includegraphics[width=0.98\hsize]{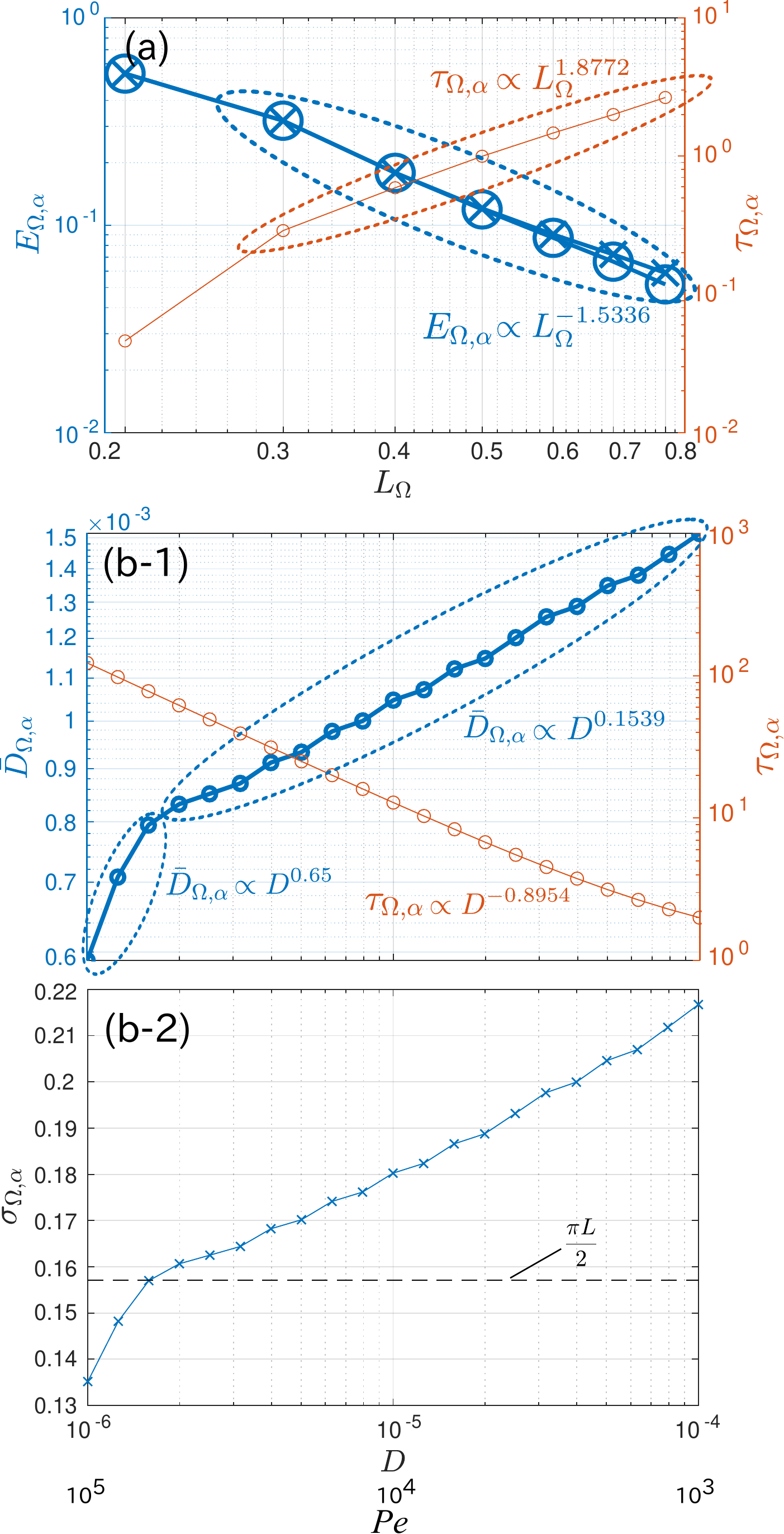}
  \caption{Simulation results of effective diffusion for the gyre flow (\ref{eq:double_gyre}):
  (a) $L_\Omega$-dependencies of $E_{\Omega,\alpha}$ (red, thin) and $\tau_{\Omega,\alpha}$ (blue, thick);
  (b-1) $D$-dependencies of $\bar{D}_{\Omega,\alpha}$ (red, thin) and $\tau_{\Omega,\alpha}$ (blue, thick); and
  (b-2) $D$-dependency of $\sigma_{\Omega,\alpha}$.}
  \label{fig:convergence_limitation}
 \end{center}
\end{figure}
Figure~\ref{fig:convergence_limitation} summarizes simulation results of the effective diffusion for the gyre flow.
In the panel~(a) of Fig.~\ref{fig:convergence_limitation}, the time constant $\tau_{\Omega,\alpha}$ and the error $E_{\Omega,\alpha}$ are shown as 
the setting \#\,1 in \tabref{tab:scale_experiment}.
The error $E_{\Omega,\alpha}$, denoted by the blue circles, becomes small as $L_\Omega$ increases,
so that the dominant phenomenon is transited from mixing and dispersion to diffusion. 
The rate of change (decay) of $E_\Omega$ is approximately $L_\Omega^{-1.5336}$,
which is faster than $L_\Omega^{-1}$ in \eqref{eq:error_convergence} derived under the scale separation.
The error $\tilde{E}_{\Omega,\alpha}$ computed by setting the initial field at $\theta_0(\vct{x})$ in \eqref{eq:initial_field}, 
denoted by blue x-marks in the panel~(a),
is in a good agreement with $E_{\Omega,\alpha}$, showing that both $\rho(\vct{x},t)$ and $\theta(\vct{x},t)$ become diffusive regardless of their different initial fields.
Thus, no consideration of the dependence on initial fields is needed to capture well the effective diffusion for the gyre flow. 
Also in the panel~(a), the time constant $\tau$ increases with $(L_\Omega/L)^{1.8772}$,
which is relevant in comparison with the well-known diffusive scaling $\tau \propto (L_\Omega/L)^{2}$ in Ref.\,\cite{multiscale_method}.

The panel~(b-1) of \figref{fig:convergence_limitation} shows the results of $\bar{D}_{\Omega,\alpha}$ and $\tau$ as the setting \#\,2 in \tabref{tab:scale_experiment}.
The rate at which $\bar{D}_{\Omega,\alpha}$ increases differs for the two regimes $D \le 10^{-5.8}$ and $D > 10^{-5.8}$; See App.~\ref{sec:transition} for the detailed discussion on the critical value $D = 10^{-5.8}$.
For $D \le 10^{-5.8}$, a linear approximation gives us an estimated rate $D^{0.65}$,
which implies the scaling law (\ref{eq:double_gyre_Deff}).
It is here noted that the estimation error from $D^{0.65}$ depends on the choice of samples used for the 
linear approximation. 
Indeed, when the four samples at $D \in \{10^{-6}, 10^{-5.9}, 10^{-5.8}, 10^{-5.7}\}$ are chosen
and a linear approximation is utilized to them,
the estimated rate becomes $D^{0.50}$.
On the other hand, the rate for $D>10^{-5.8}$ is smaller than for $D \leq 10^{-5.8}$,
showing the transition of the $\sqrt{D}$ scaling law between the two regimes.
As stated above, it suggests that the large value of the molecular diffusivity $D$ causes a breakdown of the balance between molecular diffusion and advection,
in other words, the associated boundary layer does not work in the transport phenomenon dominantly, 
i.e. $W\sub{b} \ll \sqrt{DL/U}$.

This mechanism is justified by the following observation.
For quantifying how long is the distance of movement of a fluid parcel across the periodic flow cells,
we denote by $\sigma_{\Omega,\alpha}$ the dispersion length
in the $x$- (or $y$-) direction governed by the effective diffusion as
\begin{align}
 \sigma_{\Omega,\alpha} := \sqrt{\bar{D}_{\Omega,\alpha} \tau_{\Omega,\alpha}/{n}},
\end{align}
where $n$ is the dimension of $\Omega$ and introduced in the denominator for representing the $x$- (or $y$-) directional movement
of fluid parcels. 
The panel~(b-2) of \figref{fig:convergence_limitation} shows $\sigma_{\Omega,\alpha}$ for each $D$,
where the horizontal broken line corresponds to the path length per one rotation of a circulation, $\sigma_{\Omega,\alpha} = \pi L/2$. 
Clearly, $\sigma_{\Omega,\alpha}$ is larger than $\pi L/2$ for $D > 10^{-5.8}$,
implying that a fluid parcel visits multiple cells before circulating a single cell.
The discontinuous change at $D = 10^{-5.8}$ suggests that
the fluid parcels are not trapped in the boundary layer,
and that the effective diffusion is governed by their molecular diffusion that develops over multiple cells.

\section{Scale dependence for shear flow}
\label{sec:shear_results}

In this section, we address a simple model of time-periodic shear flow
and investigate the effective diffusion arising there,
especially its scale dependence caused by a temporal oscillation in the shear.
The model flow is given in \cite{zel1982exact,isichenko1992percolation} as follows:
\begin{align}
 \vct{u}(y,t) &= \left[
 \begin{array}{c}
  U \cos(2\pi y/L) \cos(2\pi t/\tau_0)\\
  0
 \end{array}
 \right],\label{eq:shear_def}
\end{align}
where $\vct{x}= [x\ y]^\top \in \mathbb{X} = [-0.5, 0.5]\times[-0.5, 0.5]$.
The effective diffusion in the $x$-direction can be produced by the interaction of shear and molecular diffusion. 
The associated effective diffusivity is analytically determined via spatio-temporal Fourier analysis in Ref.\,\cite{majda1999simplified}.
Let $\Omega$ be the rectangle $[-0.4, 0.4]\times[-0.5, 0.5]$ with $L_\Omega = 0.4$.
Under the scale separation, which corresponds to $L \ll L_\Omega$ and $\tau_0 \ll \tau$ in Ref.\,\cite{multiscale_method}, 
the effective diffusivity in the $x$-direction, denoted by $\bar{D}_{xx}$, is described with the following analytic formula
\cite{zel1982exact,isichenko1992percolation,majda1999simplified}:
\begin{align}
 \bar{D}_{xx} &= D + \frac{D}{2}\frac{U^2}{(L/\tau_0)^2 + (2\pi D/L)^2}.
 \label{eq:shear_Deff_def}
\end{align}
This leads to the associated PDE for the effective diffusion as 
\begin{align}
 \partial_t \bar{\theta}(\vct{x},t) &= (\bar{D}_{xx} \partial_x^2 + D\partial_y^2)\bar{\theta}(\vct{x},t),
 \label{eq:shear_Deff_matrix}
\end{align}
where $\partial_x$ and $\partial_y$ stand for the differential operators in $x$ and $y$. 

\begin{figure}[!tb]
 \begin{center}
  \includegraphics[width=0.9\hsize]{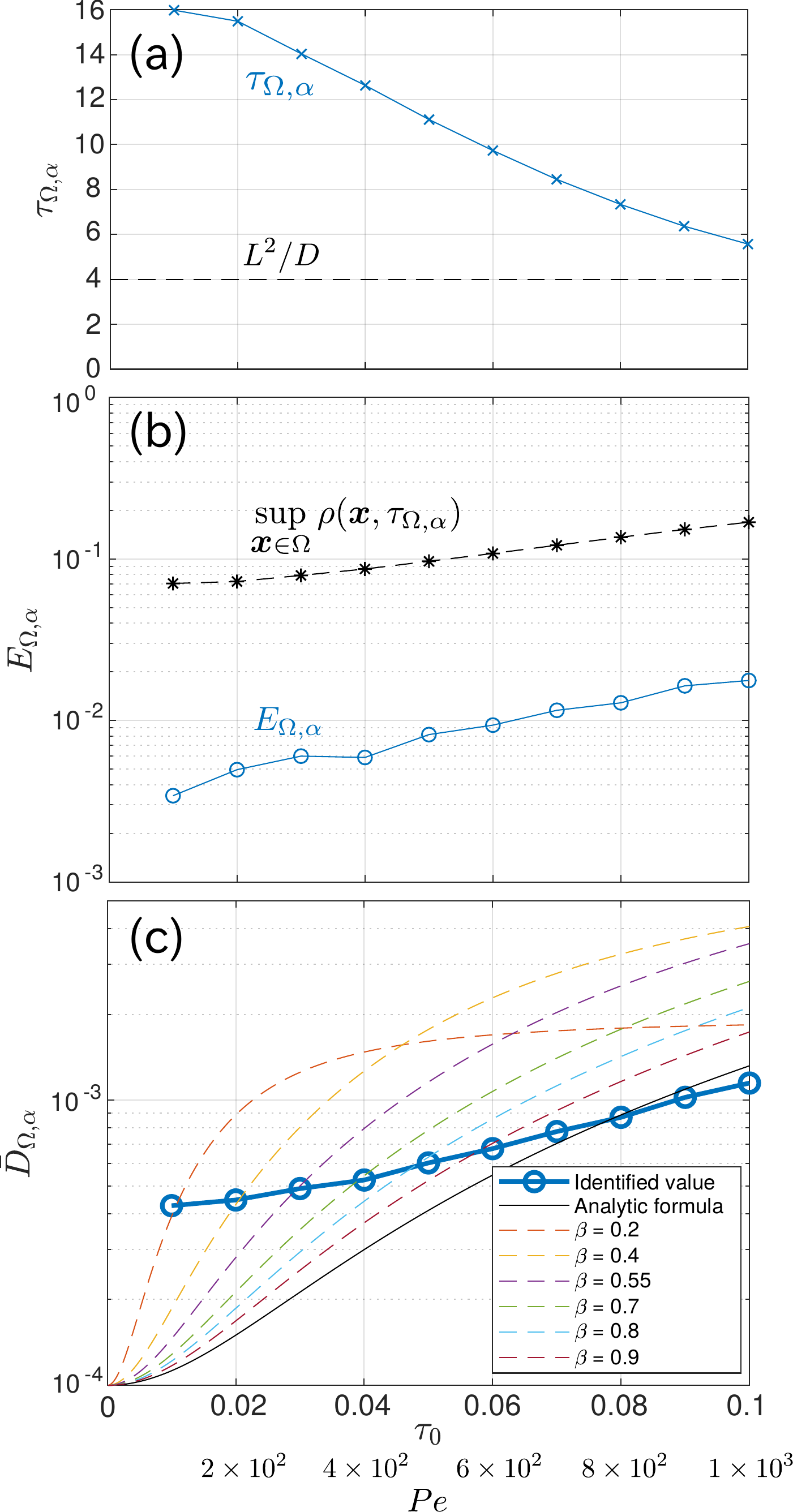}
  \caption{Simulation results of effective diffusion for the shear flow (\ref{eq:shear_def}):
  (a) $\tau_{\Omega,\alpha}$ (solid)
  (b) $E_{\Omega,\alpha}$ (solid) and $\sup_{\vct{x}\in\Omega}\rho(\vct{x},\tau_{\Omega,\alpha})$ (broken);
  (c) $\bar{D}_{\Omega,\alpha}$ (blue, solid),
  $\bar{D}_{xx}$ (black, solid) and $\bar{D}_{\beta}$ (broken)
  for each time-period $\tau_0$.}
  \label{fig:shear_results}
 \end{center}
\end{figure}

To clarify the scale dependence to be studied here,
we explain the kinematic origin of the analytic formula (\ref{eq:shear_Deff_def})
based on Ref.\,\cite{majda1999simplified}. 
Without the presence of molecular diffusion, a fluid parcel would move along the streamlines of the shear,
which are straight lines parallel to the $x$-axis,
at a ballistic rate (implying that the distance for the movement grows linearly in time).
The presence of molecular diffusion enables the parcel to move from its initial streamline onto others with velocities in the opposite direction of the initial one, 
thereby suppressing the ballistic motion and making a diffusive transport dominant instead.
In contrast, the temporal oscillations in the shear induce bounded oscillations of fluid parcels
rather than the unidirectional ballistic motion,
therefore disturbing the transport of parcels.
Thus, the effective diffusivity $\bar{D}_{xx}$ depends on $D$ and decays as the period $\tau_0$ decreases;
see \eqref{eq:shear_Deff_def} and the black line in \figref{fig:shear_results}(c) for details.
The dependence of $\bar{D}_{xx}$ on not only $D$ but also $\tau_0$ is crucial to our current study.

The scale dependence which we will study is related to the deviation of the estimated effective diffusivity from the analytic formula (\ref{eq:shear_Deff_def}).
Regarding this, it should be emphasized that the mechanism explained above
implicitly relies on the assumption of scale separation; namely,
$\tau$ should be so large that a fluid parcel can diffusively move between different streamlines in the shear.
In this, the diffusive behavior of fluid parcels has not been clearly investigated for a small timescale $\tau$.
By taking a large value of the period $\tau_0$ of the shear,
$\tau$ can be small such that fluid parcels can reach the boundary $\partial\Omega$ by the ballistic motion
before moving between the streamlines diffusively. 
For such a small $\tau$, we will estimate the effective diffusivity
and investigate how the estimated value is affected by the dominance of the ballistic motion.
Below, in order to consider not only $D$ but also multiple values of $\tau_0$, 
we regard as the length-scale not $L$ (the interval between the streamlines)
but $U\tau_0$ (the length of the ballistic movement) and rewrite the P\'{e}clet number as the leading parameter like
\begin{align}
 Pe := \frac{U^2 \tau_0}{D}.
 \label{eq:new_peclet_num}
\end{align}

Let us summarize the setting of numerical simulations.
We used $\sigma = 0.04$, $U=1$, $L = 0.02$, and $D=10^{-4}$,
and varied $\tau_0$ through $0.01, 0.02, \ldots, 0.1$
so as to sweep the advective time-scale while fixing the diffusive time-scale.
The parameter $\alpha$ was set at 0.05 as in \secref{sec:gyre_results}. 
To explicitly search $\bar{D}_{xx}$, the diffusion operator  $\bar{D}\Delta$ was modified into $\bar{D}\partial_x^2 + D\partial_y^2$
in Eqs.~(\ref{eq:Deff_eq}), (\ref{eq:test_meanflow_diff}), and (\ref{eq:test_diff}).
The simulation and optimization schemes were the same as in \secref{sec:gyre_results}, 
and the discretization steps were set at 0.002 in space and at 0.001 in time.

Figure~\ref{fig:shear_results} summarizes simulation results of the effective diffusion for the shear flow. 
The P\'eclet number $Pe$ is denoted for each $\tau_0$ at the bottom of \figref{fig:shear_results}.
In the panel~(a), the time-scale $\tau_{\Omega,\alpha}$ is shown with the blue line
and takes small values as $\tau_0$ increases.
This implies that a fast oscillatory component in the shear makes the length of the ballistic motion $U\tau_0$
small so that few fluid parcels can reach the boundaries $x = \pm 0.4$.
Also, $\tau_{\Omega,\alpha}$ has a value with the same order as $L^2/D = 4$,
which represents the diffusive time-constant in the $y$-direction.
Thus, $\tau_{\Omega,\alpha}$ becomes so small that the ballistic motion is dominant
by comparison with the diffusive movement between the streamlines.

The panel~(b) of Fig.~\ref{fig:shear_results} shows the error term $E_{\Omega,\alpha}$ with the blue line. 
For comparison, the $L_\infty$ norm $\sup_{\vct{x}\in\Omega}\rho(\vct{x},\tau_{\Omega,\alpha})$
is also shown with the broken line.
The order of $E_{\Omega,\alpha}$ is smaller than that of the $L_\infty$ norm. 
Thus, the solution $\bar{\rho}_{\bar{D}_{\Omega,\alpha}}(\vct{x},\tau_{\Omega,\alpha})$ of \eqref{eq:test_diff} is in a good agreement with $\rho(\vct{x},\tau_{\Omega,\alpha})$ of \eqref{eq:test_adv_diff},
showing the validity of the approximation as the effective diffusion.
Also, this implies that in the proposed method we regard 
the dispersion mainly caused by the time-periodic ballistic motion as the effective diffusion,
which is general and never clarified with the homogenization.
Here, while $E_{\Omega,\alpha}$ decreases with almost all $\tau_0$,
it takes a larger value at $\tau_0 = 0.03$ than $\tau_0 = 0.04$.
This is related to that the drift length $U\tau_0 = 0.03$ is
comparable with (or smaller than) the variance $\sigma = 0.04$
and hence that not only the bulk transport of the initial pulse $\rho_0$
but also its deformation (i.e. the geometric change from the Gaussian pulse) are emergent
and affect the validity of the approximation as the effective diffusion.
Indeed, for different values of $\sigma$,
we confirmed that $E_{\Omega,\alpha}$ took such a large value at a certain $\tau_0$ with a smaller magnitude than $\sigma/U$,
showing the effect of the above deformation.

The panel~(c) shows the effective diffusivity for each $\tau_0$ and $Pe$.
The blue circles represent the values of $\bar{D}_{\Omega,\alpha}$ identified with our method,
and the black solid line does $\bar{D}_{xx}$ computed with the analytical formula (\ref{eq:shear_Deff_def}).
As stated above, the analytic $\bar{D}_{xx}$ decays as $\tau_0$ decreases; 
however, the determined $\bar{D}_{\Omega,\alpha}$ becomes larger than $\bar{D}_{xx}$ for the small $\tau_0$.
This indicates that the ballistic motion is not well suppressed by the diffusive transport. 
To verify it, let us modify the analytical formula (\ref{eq:shear_Deff_def}) of effective diffusivity into
\begin{align}
 \tilde{D}_\beta &= D + \frac{D}{2}\frac{U^2}{(\beta L/\tau_0)^2 + (2\pi D/\beta L)^2},
 \label{eq:shear_Deff_modify}
\end{align}
where the degree of insufficiency of the diffusive transport between streamlines
is represented by shortening the diffusive length $L$ via a controllable coefficient $\beta \in (0,1)$.
In the figure we show $\tilde{D}_\beta$ for $\beta \in \{0.2, 0.4, 0.55, 0.7, 0.8, 0.9\}$ with the broken lines.
By increasing $\beta$ with $\tau_0$ (or $Pe$),
it is possible to adjust $\tilde{D}_\beta$ to the identified $\bar{D}_{\Omega,\alpha}$ for each $\tau_0$. 
This implies that
a fast oscillation in the shear suppresses the unidirectional ballistic motion instead of the molecular diffusion, 
and that the effective diffusion for a finite magnitude of $\tau_0$ is mainly governed by the shear.

\section{Concluding remarks}
\label{sec:conclusion}

This paper is devoted to the scale dependence of the effective diffusivity $D\sub{eff}$ for the two rudimentary flow models. 
Technically, we investigated how the $Pe$ dependence of $D\sub{eff}$ was affected
by the parameters $L_\Omega$ and $\tau$ with finite magnitudes.
To do this, for given $L_\Omega$, we developed a pulse-based method for identifying $\tau$ and $D\sub{eff}$
based on finite-time evolution of the Gaussian pulse function in \secref{sec:adv_diff_effective_diff}.
For the time-invariant gyre flow in \secref{sec:gyre_results},
the proposed method successfully reproduced the well-known error convergence and scaling law,
showing its validity.
Also, by enhancing the molecular diffusivity $D$ for the gyre flow,
we show that the scaling exponent of $D\sub{eff}$ can change according to
the breakdown of a balance between advection and diffusion in a single flow cell.
For the time-periodic shear flow in \secref{sec:shear_results},
we show that the effective diffusivity $D\sub{eff}$ can deviate from
the Fourier-based analytic formula (\ref{eq:shear_Deff_def}),
which has not been clearly reported in literature to the best of our survey.
We point out that the diffusive transport between streamlines can be insufficient in a finite time-scale of the effective diffusion for the shear,
and hence the deviation from \eqref{eq:shear_Deff_def} originates from the suppression of a ballistic motion due to the temporal oscillation in the shear.

Here, we revisit the kinematic origins of the scale dependence of effective diffusivity delineated in the two models.
The origins are commonly related to how fluid parcels are trapped in the flow structures, i.e., flow cells of the gyre in \secref{sec:gyre_results} and streamlines of the shear in \secref{sec:shear_results}.
It is general beyond the two models and central to the research on mixing and dispersion by fluid flows \cite{ottino1989kinematics}.
To show this concretely, we refer to Ref.\,\cite{tang2012finite}
that investigates the effective diffusion for a time-dependent shear model different from \eqref{eq:shear_def}.
In this, as a metric of the effective diffusion,
the authors computed the {\em variance} of fluid parcels starting from a certain initial position. 
Then, by varying the initial position, they 
illustrated the spatial distribution of the variances,
which showed the spatial pattern associated with the finite-time Lyapunov exponents (FTLEs) (see, e.g., Ref.\,\cite{haller2015lagrangian}) of the vector field.
Clearly, the formation of such a spatial pattern depends on how fluid parcels are trapped in the flow structures described by FTLE,
showing the similarity with what we showed in this paper.
Also, the authors of Ref.\,\cite{tang2012finite} reported that the FTLE-induced pattern disappeared at a large value of the molecular diffusivity.
The delineated mechanism in \secref{sec:gyre_results} that a fluid parcel can move across multiple cells diffusively helps to explain the disappearing process of the pattern. 
We contend in this paper that the finite-scale effective diffusion is governed by the interplay between the fluid parcels (passive tracers) and the flow structures.

Finally, we discuss the generality of the method proposed in \secref{sec:adv_diff_effective_diff}.
The proposed method can be applied to general (non-periodic in space and time) advection-diffusion systems
if the fundamental space and time scales of a non-periodic fluid motion can be determined (that is, if $L$ and $\tau_0$ are available).
Also, by simulating the spatio-temporal evolution of fluid flows with CFD techniques
and by using it as the pre-defined field $\vct{u}(\vct{x},t)$,
the effective diffusion can be characterized even when
the concentration profile $\theta(\vct{x},t)$ affects the velocity field $\vct{u}(\vct{x},t)$
(i.e., the fluid tracers are not necessarily passive).
In addition to the computational application,
if the initial pulse $\rho_0(\vct{x})$ is realized experimentally and its evolution $\rho(\vct{x},t)$ measured (sampled), 
the proposed method makes it possible to characterize the effective diffusion of passive tracers from the measurement data on $\rho(\vct{x},t)$,
while avoiding the computational burden in terms of CFD.


\appendix
\section{Physical intuition on $D=10^{-5.8}$ in \figref{fig:convergence_limitation}(b-1)}
\label{sec:transition}

Here we draw a physical intuition on the critical value $D=10^{-5.8}$ that divides the $Pe$-dependence into the two regimes in \figref{fig:convergence_limitation}(b-1).
To do this, let us consider the intersection between the lines that illustrate two scaling laws.
For $D<10^{-5.8}$, the scaling law is given as in \eqref{eq:double_gyre_Deff}.
For $D\geq 10^{-5.8}$, we propose to understand the scaling law by modulating \eqref{eq:double_gyre_Deff} in the following manner.
Since fluid parcels are not trapped in the boundary layer
where their movement is governed by the molecular diffusion,
we enhance the nominal diffusivity $D$ into $D_{\Omega,\alpha}|_{D=10^{-5.8}}$.
Also, since the effective diffusion appears over multiple cells,
we represent its length-scale by the width $W\sub{b}|_{D=10^{-5.8}}$ between two cells
and rewrite the P\'{e}clet number as $\tilde{Pe} = W\sub{b}|_{D=10^{-5.8}}U/D$.
Then, the intersection between the lines that illustrate two scaling laws is given by
\begin{align}
 D Pe^{0.65} = D_{\Omega,\alpha} |_{D=10^{-5.8}} \tilde{Pe}^{0.1539}.
 \label{eq:Deff_balance}
\end{align}
The solution of \eqref{eq:Deff_balance} is $D = 10^{-5.8943}$.
Thus, the critical value $D =  10^{-5.8}$ corresponds to the maximum of the molecular diffusivity
such that fluid parcels are trapped in the boundary layer.

\end{document}